\documentclass{iau}
\usepackage{graphicx}
\usepackage{amsmath}
\usepackage[varg]{txfonts}
\usepackage[utf8]{inputenc}
\usepackage{siunitx}

\title[JD 11.~~3D shape of Orion\,A] %% give here short title %%
{3D shape of Orion\,A with \textit{Gaia} DR2 \\
\Large{An informed view on Star Formation Rates \\ and Efficiencies}
}

\author[Josefa E.~Gro\ss schedl]   %% give here short author list %%
{
Josefa E.~Gro\ss schedl$^1$, %% \thanks{Present address: Fluid Mech Inc., 24 The Street, Lagos, Nigeria.},
Jo\~ao Alves$^{1,2}$, 
Stefan Meingast$^1$
\and Birgit~Hasenberger$^1$
}
\affiliation{
$^1$Universit\"at Wien, Institut f\"ur Astrophysik, T\"urkenschanzstra\ss e 17, A-1180 Wien
\\ email: {\tt josefa.elisabeth.grossschedl@univie.ac.at} \\[\affilskip]
$^2$University of Vienna, Faculty of Earth Sciences, Geography and Astronomy, Data Science @ Uni Vienna
}

\pubyear{2019}
\volume{345}  %% insert here IAU Symposium No.
\jname{Origins: from the Protosun to the First Steps of Life} 
\editors{Bruce G. Elmegreen, L. Viktor T\'oth, Manuel G\"udel, eds.}

\begin{document}

\maketitle

\begin{abstract}
The giant molecular cloud Orion\,A is the closest massive star-forming region to earth ($d\sim400$\,pc). It contains the rich Orion Nebula Cluster (ONC) in the North, and low-mass star-forming regions (L1641, L1647) to the South. To get a better understanding of the differences in star formation activity, we perform an analysis of the gas mass distribution and star formation rate across the cloud. We find that the gas is roughly uniformly distributed, while, oddly, the ONC region produced about a factor of ten more stars compared to the rest of the cloud. For a better interpretation of this phenomenon, we use \textit{Gaia} DR2 parallaxes, to analyse distances of young stellar objects, using them as proxy for cloud distances. We find that the ONC region indeed lies at about 400\,pc while the low-mass star-forming parts are inclined about \SI{70}{\degree} from the plane of the sky reaching until $\sim$470\,pc. With this we estimate that Orion\,A is an about 90\,pc long filamentary cloud (about twice as long as previously assumed), with its ``Head'' (the ONC region) being ``bent'' and oriented towards the galactic mid-plane. This striking new view allows us to perform a more robust analysis of this important star-forming region in the future.
\keywords{
methods: statistical,
astrometry,
stars: distances,
stars: pre--main-sequence,
ISM: clouds
}
\end{abstract}

\firstsection % if your document starts with a section,
              % remove some space above using this command.

\section{Introduction}

The giant molecular cloud Orion\,A has been extensively studied in the past (e.g., \cite{Bally2008}), being the closest massive star-forming region to earth. The distance to the northern part of the cloud, containing the Orion Nebula Cluster (ONC), was, for example, estimated by \cite{Menten2007} to be $\sim$414\,pc. Assuming this distance for the whole cloud, leads to a projected length of about 40\,pc to 45\,pc, when using the high column-density parts of the cloud. 
The rich ONC is considered a high-mass star-forming region, while only very few of the most massive stars in the cluster's center produce the prominent Orion Nebula (e.g., \cite{Odell2008}). The southern parts of the cloud (L1641 and L1647) produce predominantly low-mass stars, comparable to other nearby star-forming regions. For simplicity we call the ONC region the ``Head'', and the low-mass star-forming parts the ``Tail'' of the cloud. The difference in star formation activity in these two parts of the cloud is a well known fact, while we still do not fully understand the origin of the observed phenomenon. To get a better understanding of the cloud properties, we investigate the clouds gas content, star formation rate (SFR), and efficiency (SFE) in Sect.~\ref{sfr}, and the distances of young stellar objects (YSOs) using \textit{Gaia} DR2 parallaxes in Sect.~\ref{gaia}.

\section{Mass distribution and star formation rate and efficiency}  \label{sfr}

\begin{figure}[ht!]
	\begin{center}
	\includegraphics[width=0.85\textwidth]{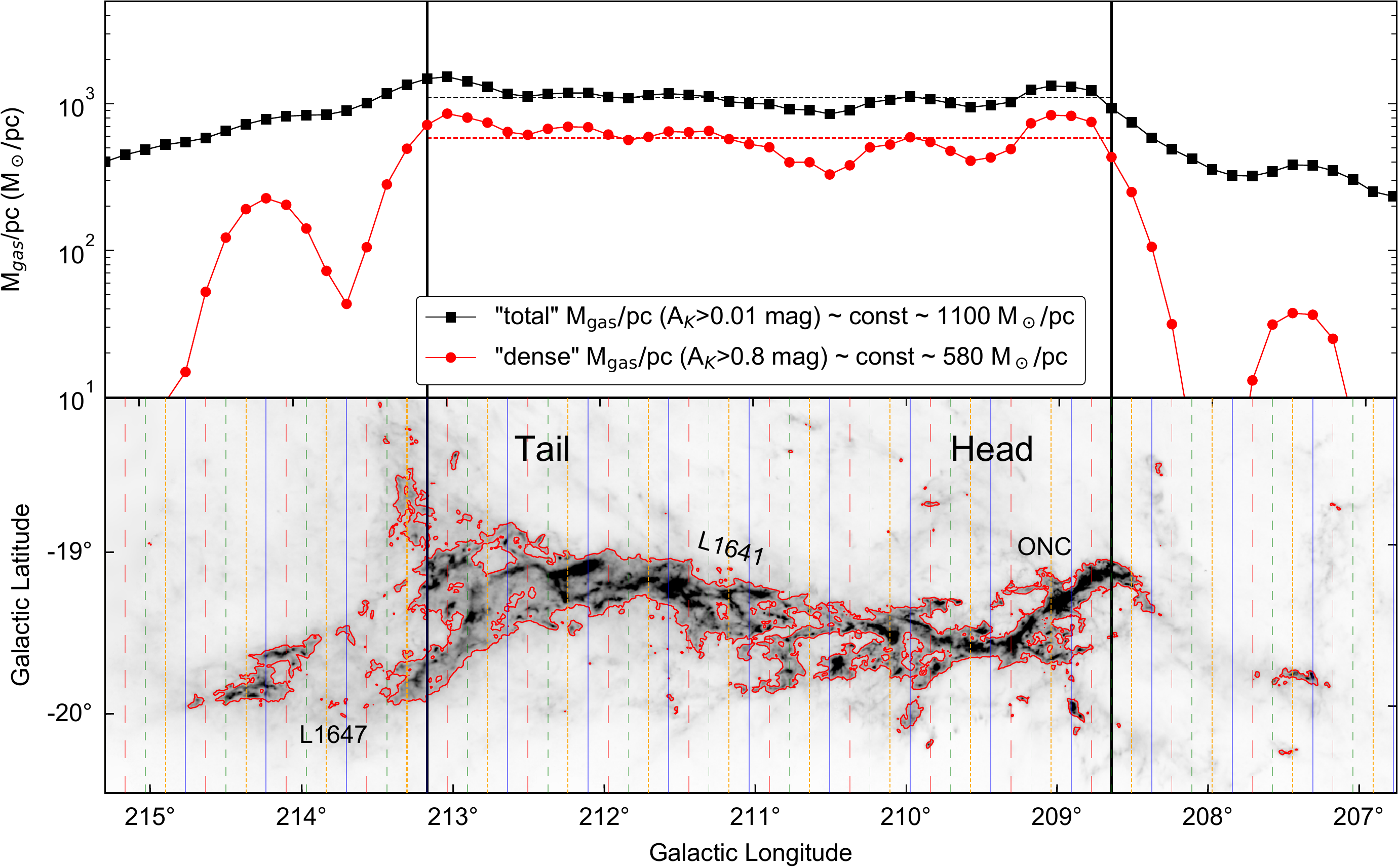}
	\caption{Gas mass distribution across the Orion\,A cloud. The cloud mass is calculated along $l$ per \SI{0.5}{\degree} wide bins ($\sim$3.6\,pc at 414\,pc, factor 4 over-sampled), and is give in the top panel in M$_\odot$/pc. The bins are indicated by the vertical lines in the bottom map, while a data point in the top figure is centred above the corresponding bin in the map (e.g., between two dashed lines). Shown separately are the ``total'' and ``dense'' gas mass distributions along the cloud. The first is evaluated for pixels with $A_\mathrm{K}>0.01$\,mag, which are basically all pixels in the displayed area, and the second is outlined by the contours at $A_\mathrm{K}=0.8$\,mag. The two vertical solid lines in both panels indicate the cloud boundaries, to exclude regions without significant gas mass, while in Galactic latitudes the bins are restricted by the image boundaries ($\SI{208.6}{\degree} < l < \SI{213.2}{\degree}$, $\SI{-20.5}{\degree} < b < \SI{-18.1}{\degree}$). The two horizontal dashed lines in the top panel show the mean M$_\odot$/pc within the boundaries, as given in the legend.}
	\label{fig:gas}
	\end{center}
\end{figure}

At first guess, one might assume that there is more gas mass available at the ONC region that fuels the relatively high star formation rate. However, when investigating the gas mass distribution across the cloud, we find that it is approximately constant (along Galactic longitudes $l$). This is highlighted in Fig.~\ref{fig:gas}, where we calculated the gas mass using the dust column-density from a Herschel-Planck-Extinction map (\cite{Lombardi2014}). For this analysis, we separately look at the gas content within the whole analyzed region (``total'') and within higher column-density regions (``dense'')\footnote{We do not know the true volume density, but only column-densities, therefore ``dense'' within parentheses. The ''total'' gas mass is restricted to the displayed region in Fig.~\ref{fig:gas}.}. The latter we define by a star-formation threshold of $A_\mathrm{K} > \SI{0.8}{mag}$, as suggested by \cite[Lada {\it et al.} (2010)]{Lada2010}. 
Extinction $A_\mathrm{K}$ is derived from optical depth by $A_\mathrm{K} \mathrm{(mag)} = \tau \cdot 3050$ (\cite{Meingast2018}), and then converted to solar masses for each pixel as follows: 
\begin{equation} \label{equ:convfactor}
M_\mathrm{gas} = A_\mathrm{K} \times ( N_\mathrm{H}/A_\mathrm{K} \cdot M_\mathrm{p} \cdot \mu_\mathrm{He} \cdot a_\mathrm{pix} )
\end{equation}
with $N_\mathrm{H}/A_\mathrm{K} = 1.24 \times 10^{22}$ cm$^{-2}$ mag$^{-1}$ (\cite{Hasenberger2016}), proton mass $M_\mathrm{p} = 1.67 \times 10^{-27}$\,kg, correction for 10\% He abundance $\mu_\mathrm{He} = 1.37$, and the pixel area $a_\mathrm{pix} = 8.63\times10^{33} \mathrm{cm}^2$ (assuming $d=414$\,pc for each $15''$ sized Herschel-map pixel). To highlight variations along the cloud we split the cloud into evenly sized bins along $l$ with $\Delta l = \SI{0.5}{\degree}$ ($\sim$3.6\,pc at 414\,pc) and find that the distribution of the gas mass is roughly constant for both the ``total'' and ``dense'' gas mass distributions within the cloud boundaries (Gro\ss schedl \textit{et al.}~in prep.).

\begin{figure}[ht!]
	\begin{center}
	 \includegraphics[width=0.85\textwidth]{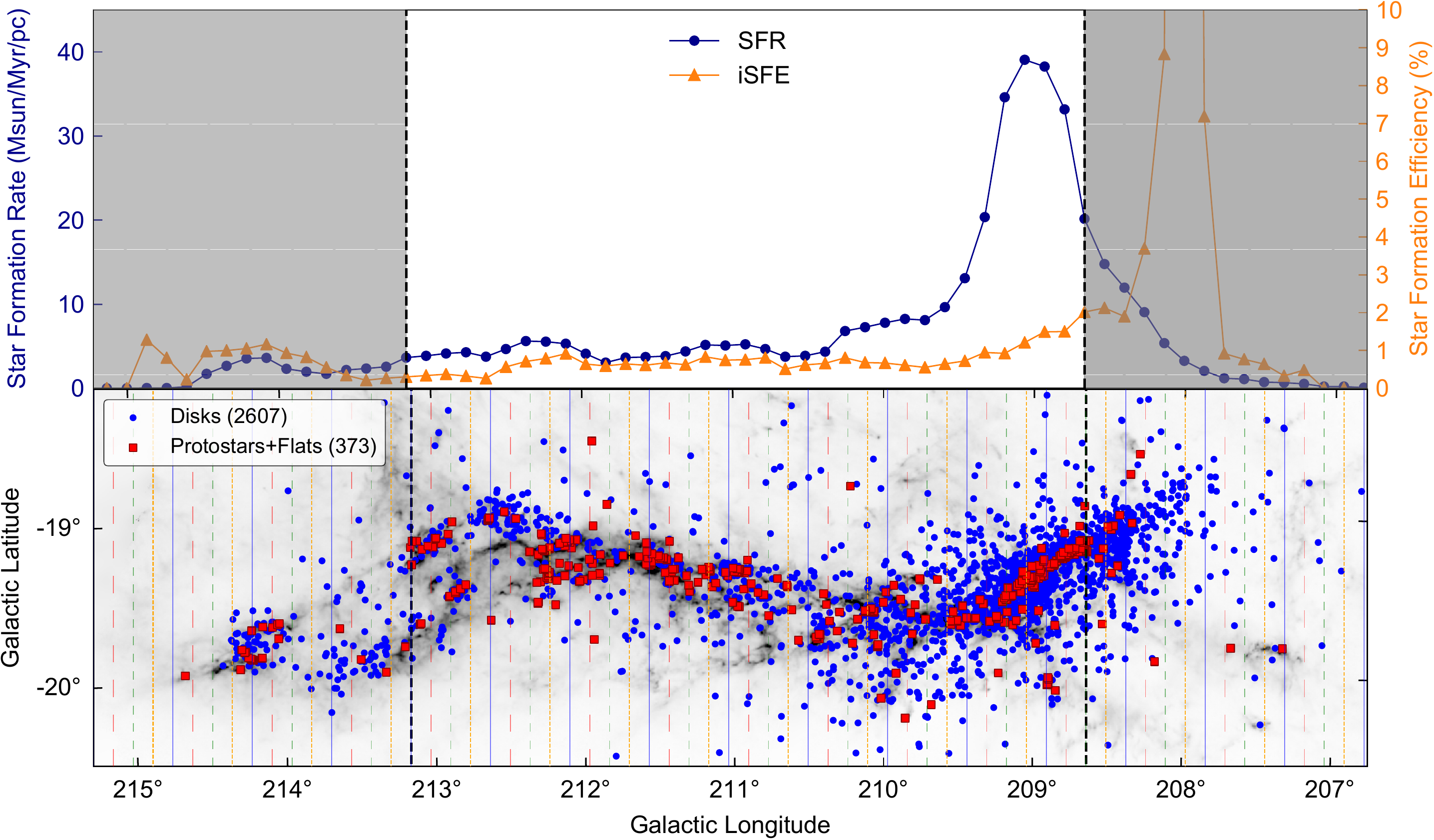}
 	\caption{Star formation rate and efficiency across the cloud. The distribution of SFR and instantaneous SFE is calculated per bin across the cloud, similar to Fig.~\ref{fig:gas}. Regions beyond the obvious cloud boundaries should be ignored (gray shaded area, top panel), since they lack significant gas mass (see also Fig.~\ref{fig:gas}). Note the different axes labelling for SFR (left) and SFE (right). The whole YSO sample is used to calculate the SFR, and the younger protostars+flats are used to evaluate the SFE in combination with the gas mass within $A_\mathrm{K}>0.8$\,mag.}
   	\label{fig:sfr_sfe}
	\end{center}
\end{figure}

This seems to be at odds with the observed higher SFR at the Head of the cloud. 
To get a more quantitative estimate of the SFR, we show its distribution across the cloud (Fig.~\ref{fig:sfr_sfe}) in a similar manner as the gas mass distribution. To this end we use the YSO catalog of \cite{Grossschedl2018b}, which contains a clean sample of 2980 IR-excess YSOs for the whole Orion\,A region. The 2D distribution of these YSOs is shown in Galactic coordinates in Fig.~\ref{fig:sfr_sfe} (bottom), displayed separately for the more evolved Class\,II YSOs (2607 disks) and the younger Class\,0/I and flat-spectrum sources (373 protostars+flats).
The SFR per parsec is calculated as follows, using the whole YSO sample:
\begin{equation} \label{equ:sfr}
\mathrm{SFR} \, (M_{\odot}/\mathrm{Myr} /\mathrm{pc}) =  N_\mathrm{YSO} \cdot \bar{M}_\mathrm{IMF} / t_\mathrm{YSO} / \Delta l
\end{equation}
with the mean IMF mass $\bar{M}_\mathrm{IMF} = 0.44\,M_\odot$ (using a Kroupa IMF), the average YSO lifetime $t_\mathrm{YSO} \approx 3$\,Myrs (e.g., \cite{Dunham2015}), and the bin size being $\Delta l = 3.6$\,pc, assuming a constant distance for the whole cloud.
We find that the Head of the cloud produced about a factor of ten more stars in the last 3 to 5 Myrs as compared to the Tail, while, interestingly, the current gas content of these two regions is very similar.

We further calculate the SFE by using only the younger protostars+flats sample in combination with the ``dense'' gas, since these sources are likely directly associated with the current star-forming gas. This so called instantaneous SFE (iSFE) is calculated as follows:
\begin{equation} \label{equ:sfr}
\mathrm{iSFE} \, =  N_\mathrm{P+F} \cdot \bar{M}_\mathrm{IMF} / (M_\mathrm{denseGas} + N_\mathrm{P+F} \cdot \bar{M}_\mathrm{IMF})
\end{equation}
The iSFE also shows a roughly constant distribution across the cloud, similar to the gas mass. Variations are at the order of a factor two. 

Even though the GMC Orion\,A is one of the best studied star-forming regions, it is still unclear what caused this burst of star formation at the location of the ONC. This raises the question: How well do we know this cloud?

\section{Evaluating distance variations in Orion\,A using \textit{Gaia} DR2}  \label{gaia}

\begin{figure}[ht!]
	\begin{center}
	 \includegraphics[width=0.85\textwidth]{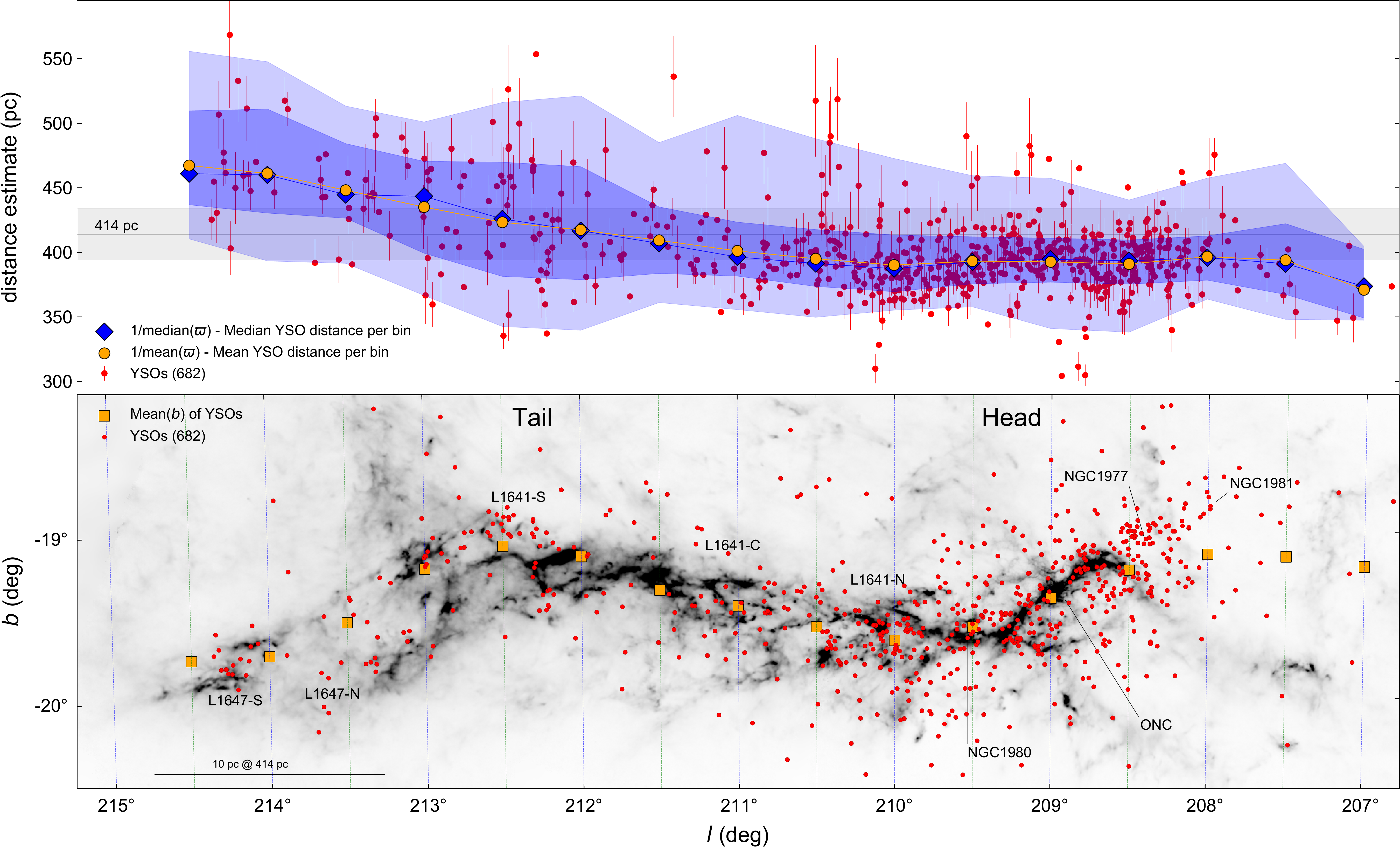}
 	\caption{Average distances of IR-YSOs along $l$. The YSOs parallaxes are averaged within $\Delta l = \SI{1}{\degree}$ wide bins (factor 2 over-sampled), shown by the circles (mean) and the diamonds (median) in the top panel. The shaded areas in the top pannel show the $1\sigma$ and $2\sigma$ confidence intervals.}
   	\label{fig:mean}
	\end{center}
\end{figure}

\begin{figure}
  \begin{minipage}[c]{0.42\textwidth}
    \includegraphics[width=\textwidth]{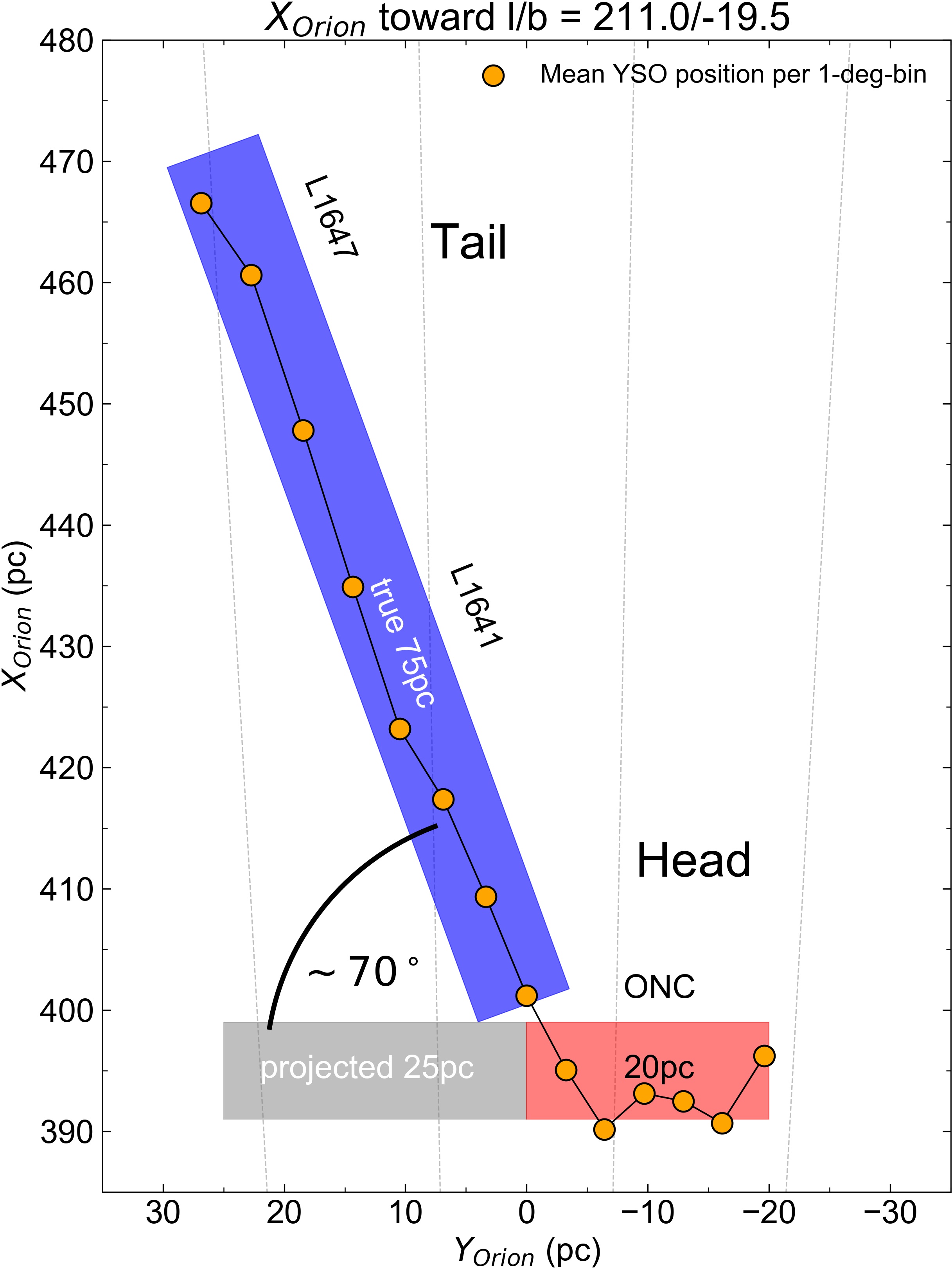}
  \end{minipage}\hfill
  \begin{minipage}[c]{0.54\textwidth}
    \caption{
       Orientation of the Orion\,A cloud displayed in a Galactic cartesian coordinate system, with X$_\mathrm{Orion}$ pointing toward Orion\,A, at $(l, b) = (211.0, -19.5)$. The sun is at (0,0), Y$_\mathrm{Orion}$ is similar to $l$, and Z$_\mathrm{Orion}$ is similar to $b$ (Z is not shown here). The circles are the averaged YSO distances as derived in Fig.~\ref{fig:mean}. The faint dashed lines are lines of constant $l$. We find that the Head of the cloud (ONC region) lies almost parallel to the plane of the sky at about 400\,pc, while the Tail is inclined about \SI{70}{\degree} away from the plane of the sky. This leads to an $\sim$75\,pc long tail instead of the projected $\sim$25\,pc. This makes the Orion\,A cloud an about 90\,pc long filamentary structure, which is twice as long as when assuming a constant distance for the whole region.
    } \label{fig:xy_orion}
  \end{minipage}
\end{figure}

\begin{figure}[h!]
	\begin{center}
	 \includegraphics[width=0.99\textwidth]{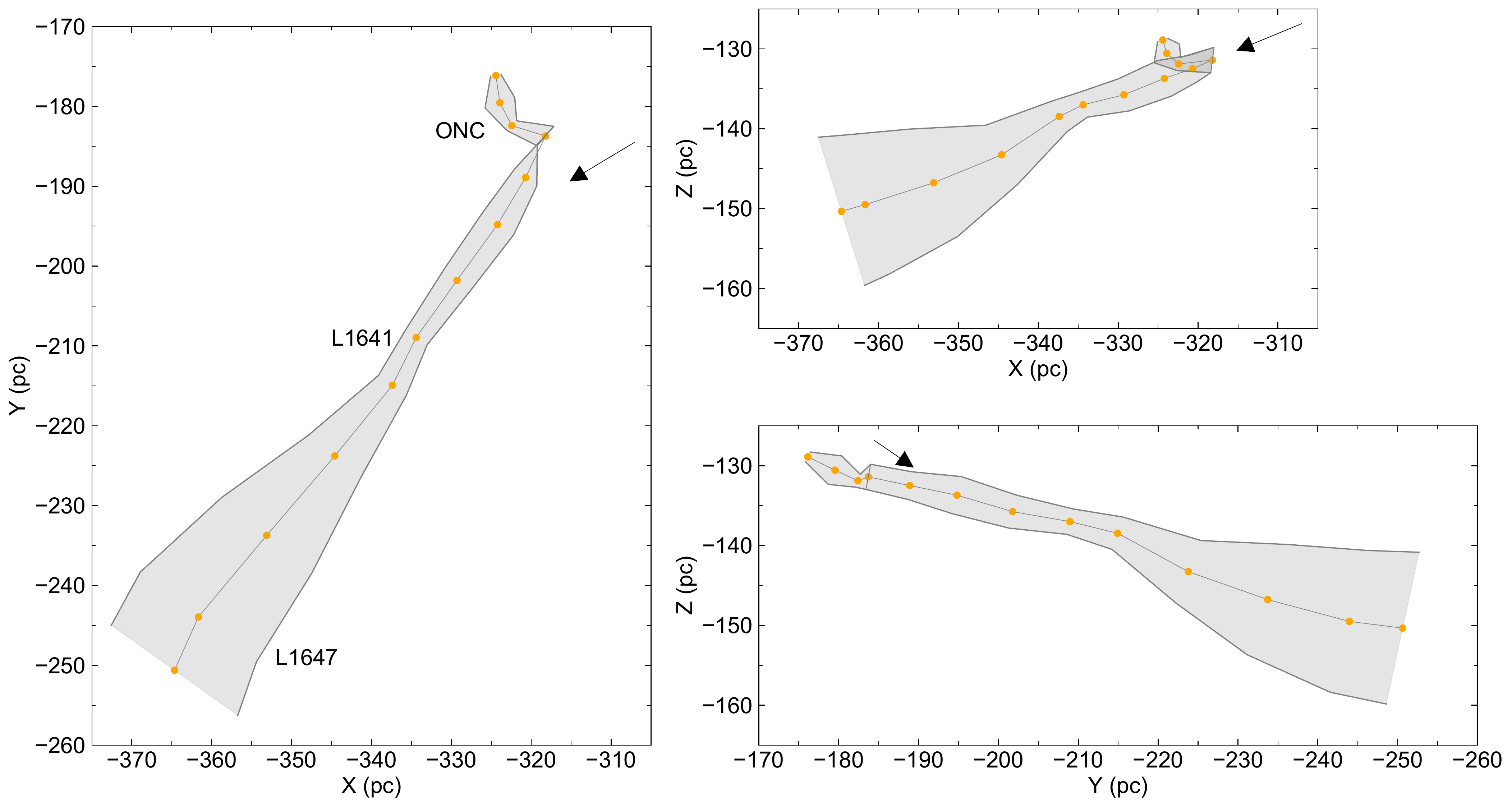}
 	\caption{3D orientation of the Orion\,A cloud displayed in Galactic cartesian coordinates. 
	The Sun is at (0,0,0), the X-axis is positive toward the galactic center, the Y-axis is positive toward the direction of Galactic rotation, and the Z-axis is positive toward the Galactic North. The dots are the averaged YSO distances as derived in Fig.~\ref{fig:mean}. The grey shaded area shows roughly the cloud extend at $A_\mathrm{K} > 0.6$\,mag. The arrows indicate the line-of-sight from the Sun pointing toward $(l, b) = (\SI{211.0}{\degree}, \SI{-19.5}{\degree})$. 
	}
   	\label{fig:3d}
	\end{center}
\end{figure}

To get potential new insights into this prominent region, we use recent \textit{Gaia} DR2 data (e.g., \cite{Prusti2016, Brown2018b}) to look for distance variations of the IR-excess YSOs connected to Orion\,A. Due to the youth of these sources they are likely still spatially connected to the cloud regions out of which they formed, or should not have moved too far from these sites. Together with the fact, that they share the same radial velocity as the gas (e.g., \cite{Hacar2016}), it makes them on average a good proxy for cloud distances. With this, we can estimate the cloud 3D shape and orientation for the first time. The following methods and results are discussed in detail in \cite{Grossschedl2018a}.
After cross-matching the YSO catalog from \cite{Grossschedl2018b} with \textit{Gaia} DR2 parallaxes, we end up with roughly 700 IR-YSOs in this region. By averaging YSO distances across the cloud (see Fig.~\ref{fig:mean}), we find that the far end of the Tail is about 70\,pc more distant than the Head (ONC), leading to a total length of the cloud of about 90\,pc (see Fig.~\ref{fig:xy_orion}). This doubles the previously assumed length of this benchmark cloud. 
Putting the averaged distances in Galactic cartesian coordinates (Fig.~\ref{fig:3d}) with X pointing toward the Galactic center, we find that the Head seems to be ``bent'' with respect to the long stretched Tail. Overall, the cloud has a cometary shape pointing toward the Galactic mid-plane.

\section{Conclusions and Implications}

This striking new view of Orion\,A will have an impact on some of the observables in this region, like cloud mass, stellar masses, luminosities, or binary separations. The fact that the Tail is differently oriented and more stretched than previously assumed, also calls for a revision of the gas mass distribution and SFR per parsec. This new ``image'' of the cloud further allows one to speculate on its formation mechanisms. 

For example, the 90\,pc estimated length of Orion\,A makes it by far the largest GMC in the solar neighborhood. Together with its aspect ratio ($\sim$30:1) and dense gas mass fraction ($\sim$45\%), the Orion\,A cloud is similar to large-scale Galactic filaments, or so-called ``bones'' of the Milky Way, as discussed in \cite{Zucker2018}. However, Orion\,A lies almost an order of magnitude farther from the Galactic mid-plane as typically found for these structures (about 100 to 130\,pc below the plane, assuming $Z_\mathrm{Sun} \approx +25$\,pc). 
A common suggestion for the origin of the Orion-Monoceros complex is, that it could be the result of an impact of a high-velocity cloud from above the Galactic plane (\cite{Franco1986}). However, this seems to be at odds with the orientation of the cloud, pointing toward the Galactic mid-plane from the southern Galactic hemisphere, so more ideas are needed. 

External processes might be able to explain the shape of the cloud with the ``bent'' Head. For example, there is a foreground population near the region of the ONC (e.g., \cite{Alves2012, Bouy2014}), which could have provided the feedback necessary to bend the head. A Supernova originating from this slightly older population could have taken place in the last about 10\,Myrs to 20\,Myrs\footnote{When considering all ONC members, we find that the cluster's older members (incl.~Class\,III) have ages up to about 10\,Myrs when investigating HR-diagrams.}, triggering the burst of star formation and also shaping this region. 
Another external event could have taken place, as proposed by \cite{Fukui2018}. They suggest a cloud-cloud collision at the location of the ONC, being a trigger of star formation. If this event had taken place, it would have had also an impact on the shape of the cloud. 

Our result illustrates well the new window that is opening with \textit{Gaia} on the structure of the dense ISM. It allows to understand the 3D shape and orientation of star-forming clouds, which will refine other derived properties of these regions.

\begin{discussion}

\discuss{Kruijssen} {In star formation simulations, one side of a cloud often collapses and forms stars first, giving rise to the ``kinked'' morphology you see in Orion. Especially in a shearing potential, this can persist. Do you have any additional evidence for the idea that this is due to external influences, or could it be secular evolution?}

\discuss{Gro\ss schedl} {Yes, it could also be secular evolution. But this is something we still need to investigate, for example, by comparing the shape and orientation of Orion\,A within the Galactic disk with simulations of spiral galaxies or collapsing cylindrical clouds.}

\discuss{Khaibrakhmanov} {Is there information about kinematics of YSOs in the cloud?}

\discuss{Gro\ss schedl} {Yes, the YSOs radial velocity (RV) is known from APOGEE (e.g., \cite{Hacar2016}). From this we know that the YSOs share the same RV as the CO gas. With this information, combined with the youth of the YSOs, they are good proxies for cloud distance. So we do not think that they moved significantly far from their birth-sites. But we do not know (yet) the proper motion (PM) of the gas, which can only be measured from stars. We have a running project (VISIONS, Meingast et al.) measuring PMs of embedded YSOs in the near infrared. PMs of these sources will also give indirect information of the gas' PM for the first time.}

\end{discussion}

\end{document}